\documentclass[runningheads]{llncs}

\usepackage{fancyhdr}
\usepackage{amssymb}
\usepackage{amsmath}
\usepackage{algorithm}
\usepackage{algpseudocode}
\usepackage{graphicx}
\usepackage{subcaption}
\usepackage{tabularx}  
\usepackage{booktabs}  
\usepackage{caption}
\usepackage{cite}
\usepackage{svg}
\usepackage{hyperref}
\usepackage[misc]{ifsym}

\DeclareCaptionSubType*{algorithm}

\newcommand{\repeatthanks}{\textsuperscript{\thefootnote}}
\DeclareCaptionLabelFormat{alglabel}{Alg.~#2}
\begin{document}
\title{Federated Stain Normalization for Computational Pathology}
\titlerunning{BottleGAN}
%
\author{
  Nicolas Wagner\thanks{Equal contribution.}\inst{1}
  \and
  Moritz Fuchs\repeatthanks\inst{1} \Letter
  \and
  Yuri Tolkach\inst{2}
  \and
  Anirban Mukhopadhyay\inst{1}
}

%
\authorrunning{Wagner et al.}
%
\institute{Department of Computer Science, TU Darmstadt, Germany 
            \and
            Institute of Pathology, University Hospital Cologne, Germany}
\maketitle              
\begin{abstract}
Although deep federated learning has received much attention in recent years, progress has been made mainly in the context of natural images and barely for computational pathology. However, deep federated learning is an opportunity to create datasets that reflect the data diversity of many laboratories. Further, the effort of dataset construction can be divided among many. Unfortunately, existing algorithms cannot be easily applied to computational pathology since previous work presupposes that data distributions of laboratories must be similar. This is an unlikely assumption, mainly since different laboratories have different staining styles. As a solution, we propose BottleGAN, a generative model that can computationally align the staining styles of many laboratories and can be trained in a privacy-preserving manner to foster federated learning in computational pathology. We construct a heterogenic multi-institutional dataset based on the PESO segmentation dataset and improve the IOU by 42\% compared to existing federated learning algorithms. An implementation of BottleGAN is available at \href{https://github.com/MECLabTUDA/BottleGAN}{https://github.com/MECLabTUDA/BottleGAN}.

\end{abstract}
\section{Introduction}
Automatic processing of histological images has the potential to become an essential prerequisite for computer-assisted diagnosis, prognostication, and assessments in computational pathology (CP). If neural networks are to be used for this purpose, the standard deep learning methods need to be trained on a vast amount of labeled data. In reality, such data is hardly available in the public domain. For instance, filtered by segmentation and histology, grand-challenge.org solely offers 15 datasets. Many of these are only of limited use. Considering the number of different dyes used in histopathology and different anatomies as well as tissue structures that are examined, many more labeled datasets are necessary. This problem is exacerbated because the style of staining can differ significantly between laboratories but also within a laboratory for a variety of reasons \cite{schomig2021quality}. For instance, protocols, storage conditions, or reagents may vary. A neural network, however, should be reliable regardless of the staining style. To this end, a solution is to collect representative training data from many laboratories \cite{campanella2019clinicallarge}. Unfortunately, creating large-scale datasets is only possible with an enormous effort, both in time and money.

The concept of federated learning (FL) appears to be a solution to this as it allows distributing the dataset creation work among many and captures the data diversity of multiple laboratories. Unfortunately, existing FL algorithms either expect a publicly available representative unlabeled dataset \cite{lin2020ensemble,chang2019cronus, sattler2021fedaux} or only work if participating clients are closely aligned in their data distribution \cite{hsu2019measuring}. It is fair to assume that neither requirement is fulfilled for computational pathology. Further, previous work does not integrate unlabeled clients out of the box \cite{mcmahan2017communication, hsu2019measuring}, leading to high participation barriers.   

In this study, we propose BottleGAN, a novel generative adversarial network architecture that makes FL applicable to CP by aligning the local data distributions of laboratories through stain normalization. We pair BottleGAN with an unsupervised federated learning procedure that makes no further demands on participating clients apart from minimal hardware requirements. We demonstrate how BottleGAN can seamlessly be integrated into federated learning algorithms based on weight aggregation (WA) \cite{mcmahan2017communication, hsu2019measuring, li2018federated} for solving downstream tasks. WA algorithms train a local neural network model per client and simply average the local model weights at a server to form a global model. After WA with BottleGAN, trained models are valid for the staining styles of all laboratories that participated in the unsupervised training of BottleGAN but did not necessarily contribute annotations to WA. We demonstrate on a heterogenic multi-institutional version of the PESO \cite{bulten2019peso} dataset significant improvement through BottleGAN compared to existing work on federated learning.

\section{Related Work}
\subsection{Weight Aggregation}
Most WA algorithms are derived from the underlying idea of \textsc{FedAvg} \cite{mcmahan2017communication}. \textsc{FedAvg} assumes that each client dataset is sampled i.i.d from a common data distribution. Many improvements have been developed to handle situations in which this assumption is not entirely fulfilled (i.e. client drift) \cite{ karimireddy2020scaffold, chen2020fedbe, hsu2020federated, hsu2019measuring, reddi2020adaptive}.
The literature on federated learning in CP is very limited. Although evaluated on WSIs, Andreux et al. \cite{andreux2020siloed} use a more general adaptation of federated learning in that they learn different normalization statistics per client. Changing normalization layers has also been studied by others \cite{diao2020heterofl}. Lu et al. \cite{lu2020federatedwsi} are mainly concerned about the size of WSIs and use a pretrained model for embedding WSI patches. However, neither approach addresses the heterogeneity of participants in federated learning and goes significantly beyond standard weight aggregation algorithms. Recently, \cite{lutnick2021fedstudy} have shown in a minimal setting with only three clients that FL can create robustness of deep learning in CP to multi-institutional heterogeneity.
\subsection{Stain Normalization}
Lately, various versions of StainGAN \cite{shaban2019staingan} have become the dominant idea for stain normalization. StainGAN is an adaption of CycleGAN \cite{zhu2017cycle} and is able to learn the stain normalization task from small exemplary WSI crops while incorporating spatial variations. Unfortunately, CycleGANs, and hence StainGANs, are only able to handle one-to-one style mappings. In the FL setting under consideration in this work, the staining styles of all clients must be aligned. We are not aware of any work that is particularly intended to normalize various staining styles with only one or two neural networks. The most notable works for such many-to-many style transfers in the natural image domain, StarGAN \cite{choi2018stargan} and StarGANv2\cite{choi2020starganv2}, use a framework akin to CycleGANs but deploys a single conditional generator for all transferred styles. BottleGAN can be used as an orthogonal mechanism to improve existing WA-based FL algorithms with federated stain normalization.

\section{Method}
In the following, we first give a problem statement of FL in CP and derive the necessity of BottleGAN. Afterward, we show how BottleGAN is constructed and how it can be trained with FL.
\begin{figure}[t]
\centering
\hspace{-0.7cm}\includegraphics[width=\linewidth, height=3cm]{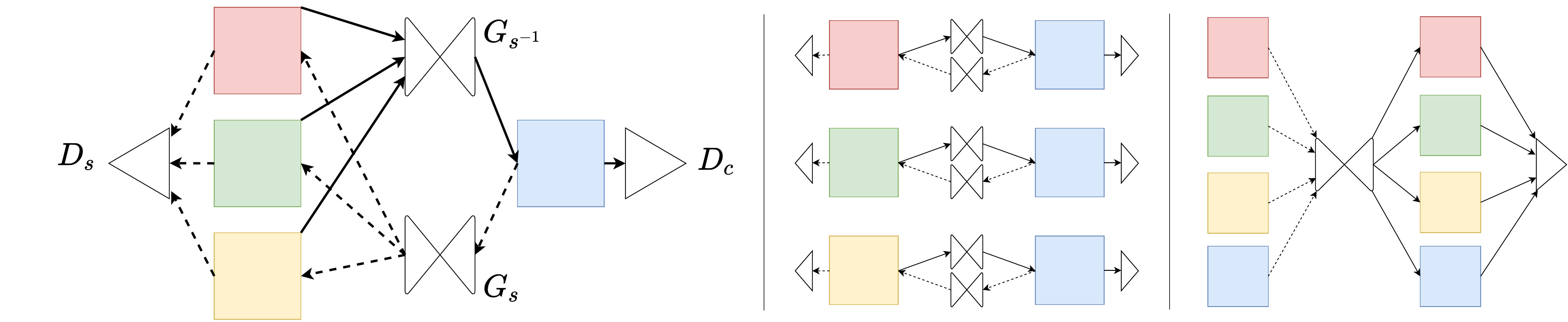}
\subfloat[\label{man} BottleGAN\newline\hspace*{-0.3cm}\textit{Many-One-Many}]{\hspace{.47\linewidth}}
\subfloat[\label{woman} StainGANs\newline\hspace*{.4cm}\textit{Many-One-One}]{\hspace{.27\linewidth}}
\subfloat[\label{diverse} StarGAN\newline\hspace*{.4cm}\textit{Many-Many}]{\hspace{.28\linewidth}}
\caption{The BottleGAN architecture trains only two generators (cones) and two discriminators (triangles) for mapping between many staining styles (squares). For many StainGANs (b), the number of neural networks is linearly dependent on the number of staining styles. The number of mappings one StarGAN (c) must represent grows quadratic with the number of staining styles whereas BottleGAN achieves linear growth.}
\label{fig:overview_Arch}
\end{figure}
\subsection{Problem Statement} In a standard deep learning setting for CP we have access to a dataset $D = \{(x_i,y_i)\}_{i=1}^{N}$ of $N$ \textit{WSI-Label} pairs. In the FL setting, however, we assume that there are clients $K$ each owning a portion $D^{k} = \{(x_i^k,y_i^k)\}_{i=1}^{N^k}$ of $D = \bigsqcup_{k=1}^K D^k$ that can not be shared with others. For a simplified notation, we omit unlabeled clients which trivially can be integrated in the following derivations. As mentioned in the introduction, the success of WA-based FL is commonly dependent on the local distributions, from which the client datasets $D^k$ were drawn, not diverging too much \cite{hsu2019measuring}. For CP, we can further decompose a WSI $x$ into a destained content image $c$ and staining style function $s$ such that $s(c) = x$.\footnote{Without loss of generality we can assume that the content image is stained in a reference staining scheme rather than destained.}
Given the staining style functions $S = \bigcup_{k=1}^{K} \{s_i^k\}_{i=1}^{N^k}$ of all clients, we define the decomposed dataset of a client as
\begin{equation}
   \mathcal{D}^k = \bigcup_{s \in S}\{(s(c^k_i), y^k_i) \}_{i=1}^{N^k}.
\end{equation}
In words, $\mathcal{D}^k$ contains the content images of a client in the staining styles of all clients. If we can create the decomposed set for each client in a privacy-preserving manner, we can also align the client data distributions much more closely.
\subsection{Network Architecture} For creating the decomposed datasets, we introduce BottleGAN, a neural network architecture that is able to map between the staining styles of all clients. The BottleGAN architecture follows a \textit{Many-One-Many} paradigm that differs conceptually from previous work as depicted in Figure \ref{fig:overview_Arch}. BottleGAN consists of two generators that perform staining style transfers and two discriminators needed for training purposes. Generator $G_{s}$ is responsible for approximating all staining style functions $s \in S$ whereas $G_{s^{-1}}$ is trained to approximate all inverse staining style functions $s^{-1}$. Hence, in contrast to a naive implementation of a \textit{Many One-One} paradigm (Figure \ref{fig:overview_Arch} (b)), i.e. one StainGAN \cite{shaban2019staingan} per staining style transfer, we are significantly more computationally efficient as the number of neural networks to train remains constant. Additionally, compared to a StarGAN-based \cite{choi2018stargan, choi2020starganv2} \textit{Many-Many} approach (Figure \ref{fig:overview_Arch} (c)), we avoid that the number of staining style transfers one network has to learn grows quadratic with the number of staining styles. Our architecture results in linear growth and, thereby, simplifies the training task considerably.
\newline\newline
\textbf{Generators} Both generators follow the same three major design choices that are novel in contrast to previous deep stain normalization networks \cite{shaban2019staingan}. 
First, we only use convolutions of size 1x1, no downsampling or upsampling techniques, and no skip connections that are usually used in common neural style transfer architectures like U-Net \cite{ronneberger2015unet} or SB-Generators \cite{karras2019sbgan}. Contrary to changing the style of a pixel in a natural image, changing the staining style of a pixel in a WSI should only depend on the pixel's content and the globally prevalent staining style. Obviously, image capturing noise and other influences weaken this assumption. Nonetheless, in the federated setting, it is desirable to keep the communication and training costs of neural networks as small as possible for clients. For CP, both can be improved if parameter-efficient networks are used that avoid modeling long-distance correlations between pixels. Another advantage is that our architecture is entirely independent of the size of the input image. Phrased differently, as most WSIs can not be processed at once by neural networks due to their size, the standard solution is moving the networks across a WSI and processing one crop after the other. In this case, architectures like U-Net probably process a pixel differently depending on its position within a crop.

Second, we follow the current success of adaptive instance normalization (AdaIN) \cite{karras2019sbgan, huang2017adain, choi2020starganv2} to condition the BottleGAN generators on a particular staining style function. Although we implement all staining style functions with one BottleGAN and all inverse functions with another, we only use one trainable style code per staining style. Gaussian noise is added to style codes to represent ambiguities in staining style functions and make BottleGAN more robust. 
\begin{figure}[t]
\begin{subfigure}{.33\textwidth}
  \centering
  \includegraphics[width=0.99\linewidth]{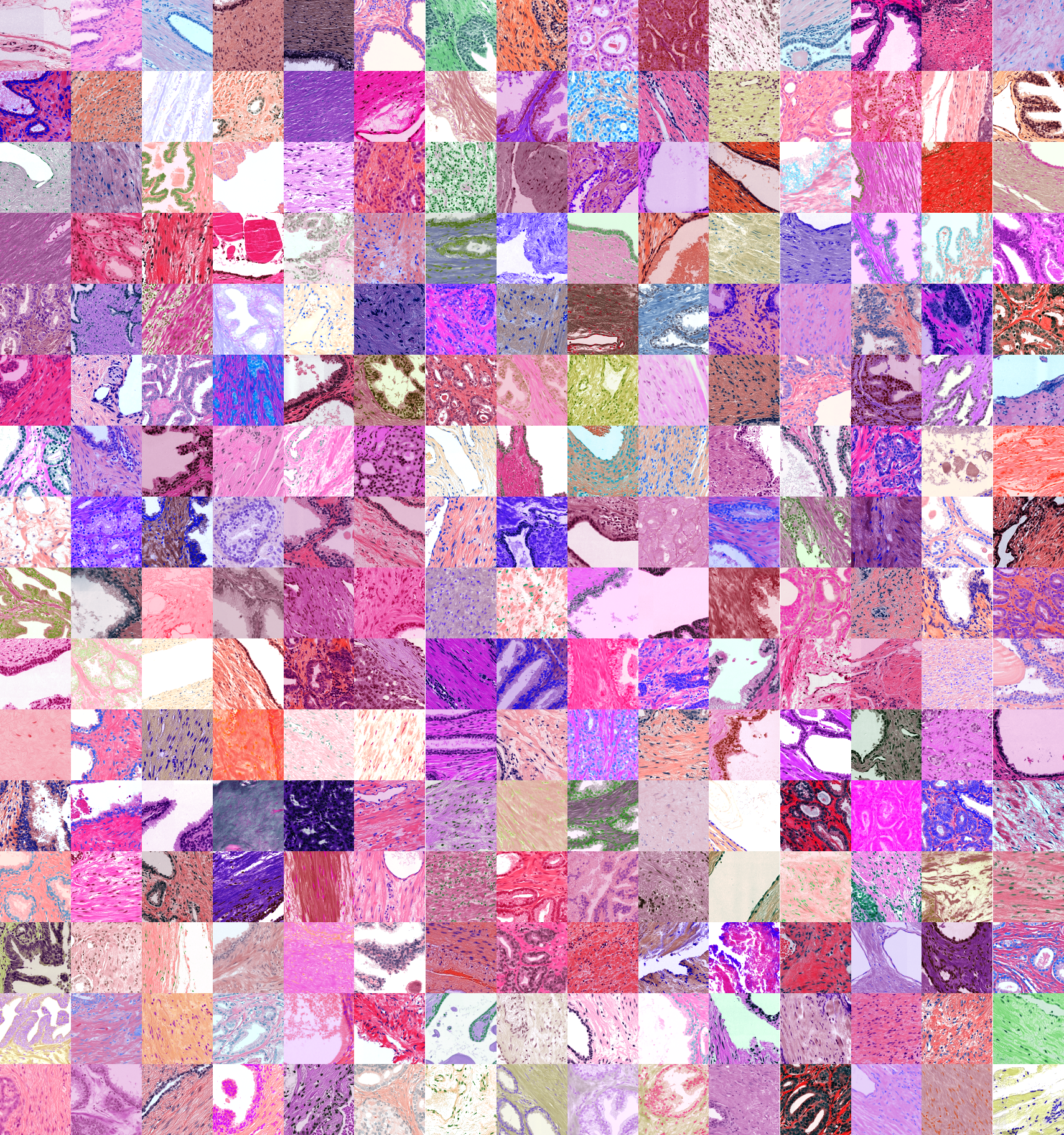}
  \caption{Input}
  \label{fig:sfig1}
\end{subfigure}%
\begin{subfigure}{.33\textwidth}
  \centering
  \includegraphics[width=0.99\linewidth]{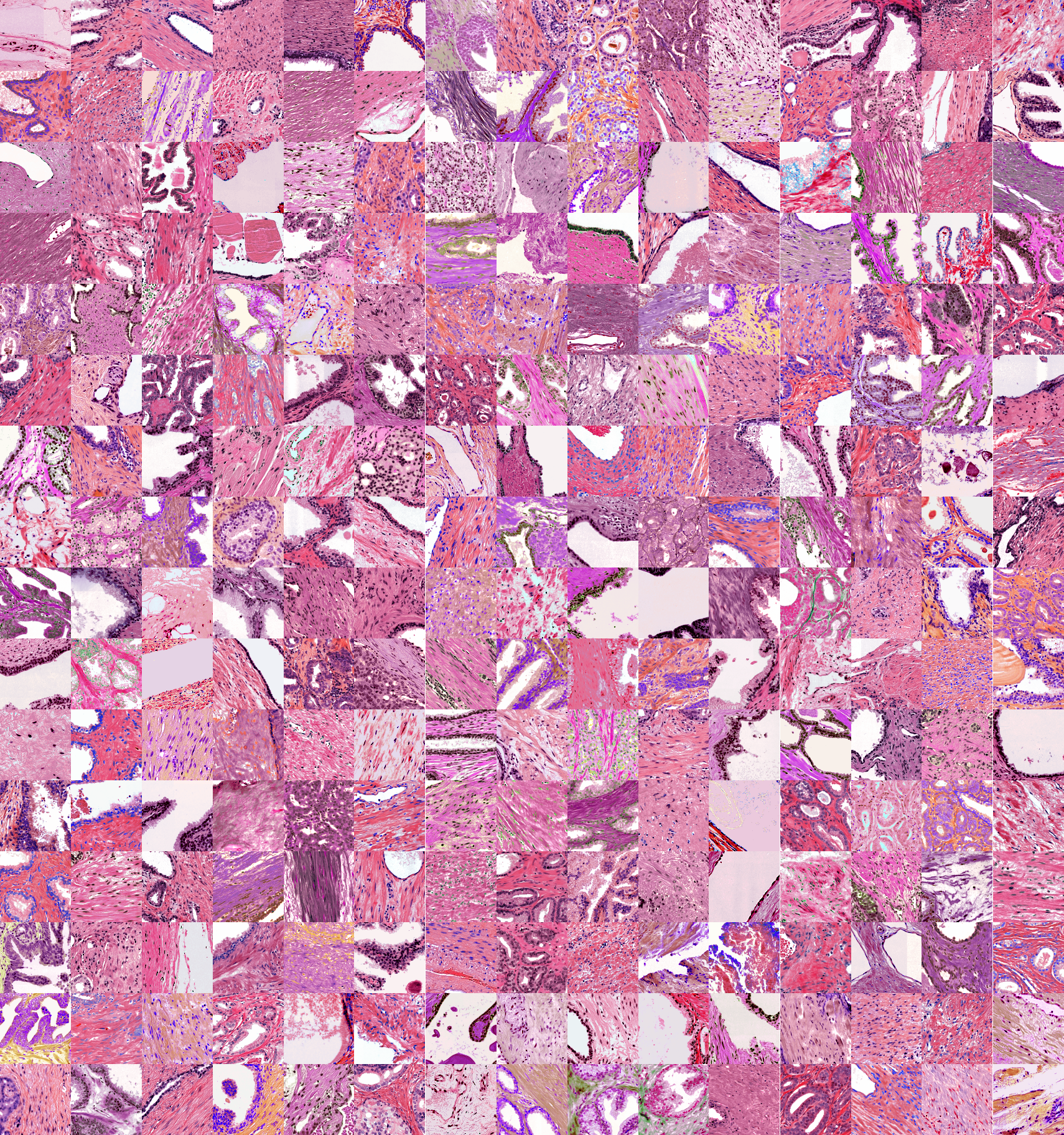}
  \caption{Normalization}
  \label{fig:sfig2}
\end{subfigure}%
\begin{subfigure}{.33\textwidth}
  \centering
  \includegraphics[ width=0.99\linewidth]{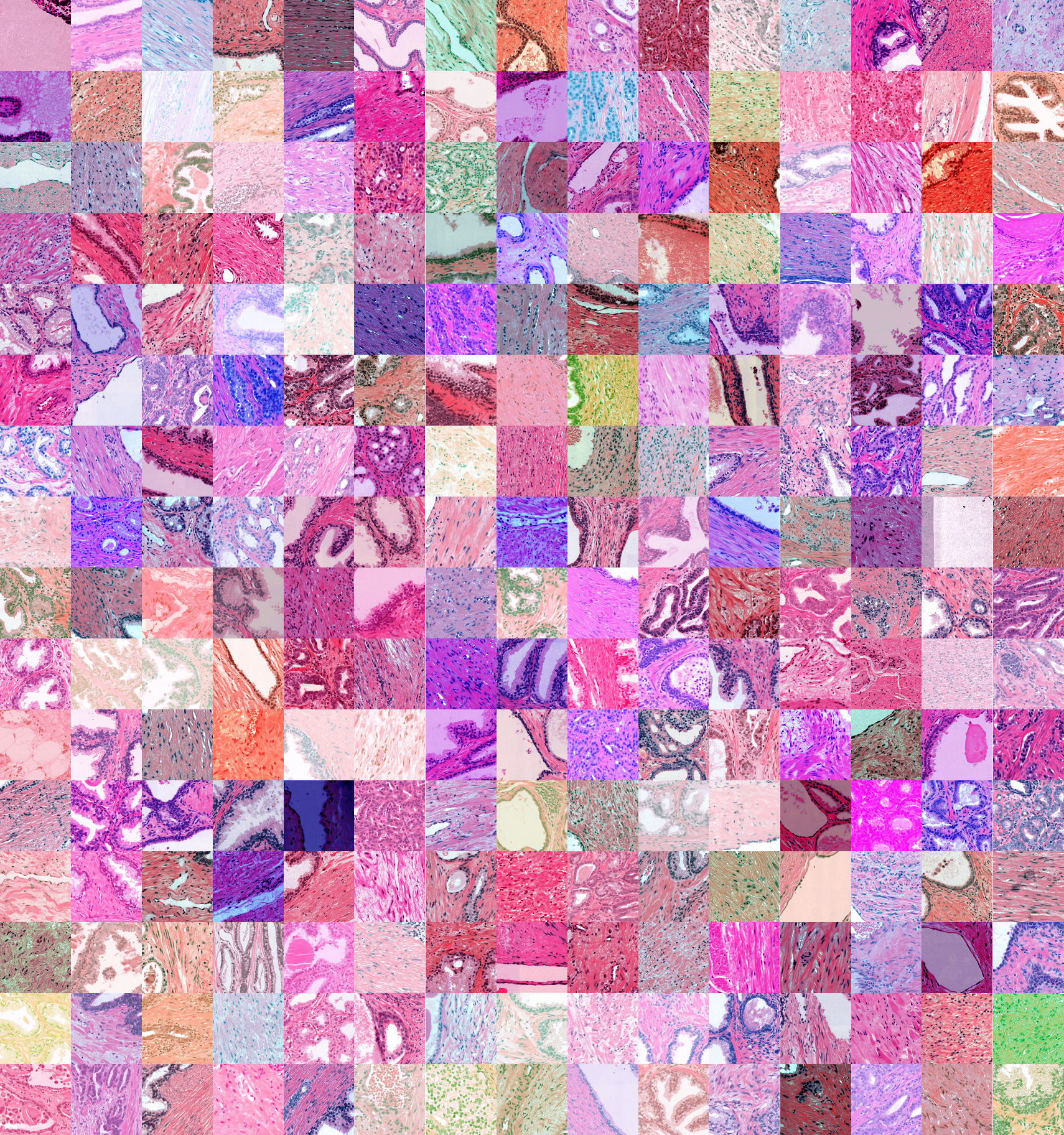}
  \caption{Restaining}
  \label{fig:sfig2}
\end{subfigure}
\caption{A visual demonstration of BottleGAN capabilities. (a) shows  240 artificial stainings targets, (b) demonstrates the normalization of (a) by BottleGAN, and (c) displays the restaining from (b) to (a) by BottleGAN.}
\label{fig:visu}
\end{figure}

Finally, both generators work directly in the optical density (OD) space, which is the negated logarithm of the image intensities. This is plausible since the staining matrix, which can describe all linear effects of staining, also acts in the OD space.
\newline\newline
\textbf{Discriminators} Looking at the discriminators now, both differ in their structure. The discriminator $D_c$ decides whether an image is destained (or reference stained). Hence, we can make use of a standard PatchGAN \cite{li2016wand} discriminator for this binary decision. The discriminator $D_s$, however, is ought to decide for all staining styles if an image is stained in a particular style or not. For this, we condition a PatchGAN discriminator by concatenating the respective style code to the output of the last downsampling convolutional layer.   
\begin{figure}[]
\centering
\includegraphics[width=1.\textwidth]{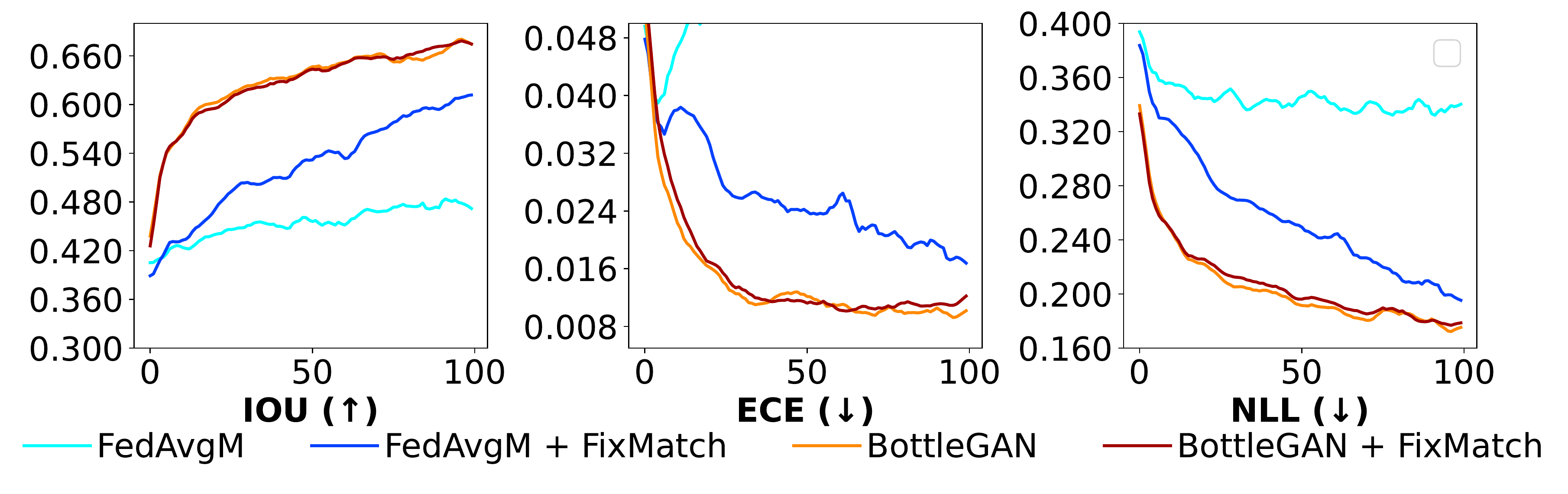}
\caption{Results on the test set of \textsc{FedAvgM} and the variants with \textsc{FixMatch} as well as BottleGAN over 100 communication rounds. \textsc{FixMatch} can seemingly improve client homogeneity but by far not to the extent of BottleGAN.}
\label{fig:pesotestwo}
\end{figure}
\subsection{Federated Learning}\label{sec:btnf}
We train BottleGAN with FL based on knowledge distillation \cite{hinton2015distilling}. Given a dataset $X = \bigsqcup_{k=1}^K X^k$ of WSIs as the union of non-shareable local datasets $X^k$ of clients $K$ and a shareable dataset $C$ of destained or reference stained WSIs owned by a server.
We start by asking all clients to train their own local BottleGAN solely between their staining style and the reference staining style defined by $C$. This requires a non-federated training similar to CycleGAN \cite{zhu2017cycle}. Afterward, the clients send their local models to the server. The server, in turn, applies the collected client generators to its own dataset $C$ to create a novel dataset $\hat{X}$ that contains C in the staining styles of all clients. The server then proceeds by training a global BottleGAN on $\hat{X}$  and $C$ as a distillation of all local BottleGANs. Finally, each client receives the global BottleGAN and is able to create the decomposed dataset. The only assumption we make is that the reference dataset $C$ is public, and this assumption seems to be mostly fulfilled based on public teaching examples alone.

An algorithmic description of the federated and the non-federated learning of BottleGAN is given in the supplementary material. Due to the enormous size of WSIs, an offline construction of the decomposed datasets might not be handy. Therefore, we state an online integration of BottleGAN into WA-based FL in the supplementary material, too.

\section{Evaluation}
\subsection{Dataset} Most available CP segmentation datasets are either too small or lack a sufficient labeling quality for the evaluation of FL algorithms. Additionally, the computational costs of simulating FL systems are massive. Hence, we limit ourselves to the PESO \cite{bulten2019peso} dataset of prostate specimens. PESO comprises 102 hematoxylin and eosin stained whole slide images (WSI) and corresponding segmentation masks.
\subsection{Experimental Setup}
\vspace{0.2cm}\hspace{0.0cm}\begin{minipage}[c]{0.45\textwidth}
\begin{tabularx}{\textwidth}{@{}p{2.4cm}lll@{}}
\toprule
\textbf{Method}                  & \textbf{IOU}            & \textbf{ECE}            & \textbf{NLL}            \\ \midrule
FedAvgM                          & 0.470          & 0.061          & 0.339          \\
+ FixMatch                           & 0.613          & 0.016          & 0.193          \\\midrule
BottleGAN                 & \textbf{0.671}          & \textbf{0.011}          & \textbf{0.177}          \\
+ FixMatch                             & 0.671          & 0.013          & 0.180              \\ \bottomrule
\end{tabularx}
\captionof{table}{The IOU {\small($\uparrow$)}, ECE {\small($\downarrow$)}, and NLL {\small($\downarrow$)} results on the test set for all evaluated methods. The homogenization of the clients through BottleGAN significantly improves all metrics.}
\end{minipage}\hfill
\begin{minipage}[c]{0.45\textwidth}
\vspace{-0.45cm}\begin{tabularx}{\textwidth}{@{}p{3.5cm}ll@{}}
\toprule
\textbf{Method}                  & \textbf{MSE}                    & \textbf{FID}            \\ \midrule
U-Net Generator                        & \textbf{0.030}                & 1.281          \\
Many-Many                         & 0.034                  & 1.385         \\
Many-One                         & 0.035                  & 1.397          \\ \midrule
BottleGAN                         & \textbf{0.030}              & \textbf{1.235}        
\textbf{}\\ \bottomrule
\end{tabularx}
\captionof{table}{Evaluated on 240 artificial staining styles, both the reconstruction MSE {\small($\downarrow$)} as well as the FID {\small($\downarrow$)} are greatly improved by BottleGAN in contrast to other architectures.}
\end{minipage}
\newline\newline
Our experiments simulate a FL setup with 20 clients for 100 training rounds. Each client owns a training and a testing WSI, respectively. Both WSIs follow the same staining style, which we establish with the Macenko algorithm. The staining styles are unique per client and are random linear combinations of the styles defined in Schömig-Markiefka et al. \cite{schomig2021quality}. To increase realism, we assume that not every client offers labels at all, and if they do, then the number of labels varies. For this purpose, we allow 60 labeled $300 \times 300$ px patches at a $10\times$ magnification among the clients, whereby half of the clients can have between 1 and 11 patches. The other half does not have ground truth annotations at all and only contributes its staining styles. We always process random image crops of size $224 \times 224$ px.

Since BottleGAN can be paired with any WA-based FL algorithm, and in line with other recent work \cite{chen2020fedbe, lin2020ensemble, jeong2020fedmatch}, we choose \textsc{FedAvgM} \cite{hsu2019measuring} as a baseline comparison. Further, as BottleGAN should be considered a federated semi-supervised learning (FSSL) algorithm due to its capability to include unlabeled clients, we also compare against the naive combination of FedAvgM and the state-of-the-art semi-supervised learning algorithm FixMatch \cite{sohn2020fixmatch}. At this, we follow the implementation of Jeong et al. \cite{jeong2020fedmatch} but stick to the naive combination as other mechanisms are orthogonal to BottleGAN. The performance of all algorithms is compared with the help of the intersection over union (IOU), the expected calibration error (ECE), and the negative-log likelihood (NLL). Results are an average over two seeds and three folds. For each fold, the labeled patches are distributed differently among the clients. We will make the implementation publicly available.

\subsection{Results}
The test results after 100 simulated communication rounds of \textsc{FedAvgM} can be read in Table 1. Further, we plot the development throughout training in Figure \ref{fig:pesotestwo}. The findings are unambiguous. The worst result is achieved in all metrics when only \textsc{FedAvgM} is applied, whereas the homogenization of client staining styles through BottleGAN seems to be a way to success for FL in CP. The addition of BottleGAN leads to significant improvements in all evaluated metrics. The IOU is increased by 0.21, the ECE is lowered by 0.50, and the NLL is reduced by 0.16. The addition of FixMatch can only improve plain \textsc{FedAvgM} without BottleGAN and even leads to slightly worse performance in terms of NLL and ECE if combined with BottleGAN. Presumably, the consistency learning of FixMatch also results in some sort of client homogenization but not to the extent BottleGAN can achieve.

\subsection{BottleGAN Architecture}
We validate major design choices of BottleGAN by training it in a non-federated setting to capture staining style transfer between 240 artificial staining styles. The staining styles are created as random linear combinations of the styles defined in Schömig-Markiefka et al. \cite{schomig2021quality}. Additionally, the entries of the corresponding stain matrices and the pixelwise optical densities are augmented with independent Gaussian noise. Exemplary artificial staining styles are displayed in Figure \ref{fig:visu} (a), the normalization of BottleGAN in Figure \ref{fig:visu} (b), and the restaining in Figure \ref{fig:visu} (c). Even for so many challenging style transfers, both the normalization and the restaining visually appear to be successfully achieved. Further experiments are evaluated by the mean squared reconstruction error (MSE), and the Fréchet inception distance (FID) \cite{heusel2017fid}. The results can be found in Table 2. First, we validate the novel generator design. BottleGAN achieves a MSE on par with the usually used U-Net \cite{ronneberger2015unet, shaban2019staingan} generator and even improves the FID while being translation invariant, using only half of the parameters, and avoiding upsampling artifacts. We also compare the novel \textit{Many-One-Many} paradigm implemented by BottleGAN to the \textit{Many-Many} paradigm of StarGAN \cite{choi2018stargan} and a \textit{Many-One} paradigm that implements BottleGAN with only one generator for both normalization and restaining. Considering the same training budget, BottleGAN greatly outperforms all other paradigms. Please note that we did not compare against the \textit{Many-One-One} paradigm, which would train many StainGANs (see Figure \ref{fig:overview_Arch} (b)) due to the intractable computational costs.

\section{Conclusion}
In this work, we introduced BottleGAN, a novel generative model to unify heterogeneous histological slides from different pathology labs in their staining style. BottleGAN is built of a new architecture tailored explicitly to staining style transfer and paired with an unsupervised FL algorithm. Further, we integrated BottleGAN into WA-based FL and demonstrated the superiority of our approach in contrast to existing FL algorithms developed for natural images. As BottleGAN allows for incorporating clients with unlabeled datasets, it becomes easier for laboratories to enter federated learning and share knowledge. In future work, we aim to incorporate uncertainty estimation into BottleGAN for building a bridge between FL and continual learning.

%
%
%
\newpage
\bibliographystyle{splncs04}
\bibliography{paper107}
\title{Supplementary Material}
\author{}
\institute{}
\maketitle
\hspace{-0.5cm}\textbf{Non-Federated Loss Functions of BottleGAN}\newline\newline
\begin{tabular}{p{65mm}p{55mm}}
\toprule\midrule
     Discriminator Loss & Generator Loss\\ \midrule
     \vbox{
    \( 
    \scriptsize
    \begin{aligned}
        \mathcal{L}_{disc} = &\frac{1}{NM} \sum_{i=1}^{N} \sum_{j=1}^{M} D_s(G_s(c_j | e_{x_i}) | e_{x_i}) \\
         & + \frac{1}{N} \sum_{i=1}^{N} D_c(G_{s^{-1}}(x_i | e_{x_i} )) \\ 
         & - \frac{1}{N} \sum_{i=1}^{N}  D_s(x_i |e_{x_i}) - \frac{1}{M} \sum_{i=j}^{M} D_c(c_j)
    \end{aligned}
    \)
    }
    & 
     \vbox{
    \(     
    \scriptsize
    \begin{aligned}
        \mathcal{L}_{gen} = & - \frac{1}{NM} \sum_{i=1}^{N} \sum_{j=1}^{M} D_s(G_s(c_j | e_{x_i}) | e_{x_i})   \\
         & - \frac{1}{N} \sum_{i=1}^{N} D_c( G_{s^{-1}}(x_i | e_{x_i} )) \\
         & + \lambda_{cyc}   \mathcal{L}_{cyc} + \lambda_{idt}   \mathcal{L}_{idt}
    \end{aligned}
    \)
     }\\\midrule
    Cycle Loss & Identity Loss\\ \midrule
     \vbox{
   \( 
       \scriptsize
        \begin{aligned}
        \mathcal{L}_{cyc} = & \frac{1}{N} \sum_{i=1}^{N} ||G_s(G_{s^{-1}}(x_i | e_{x_i} ) | e_{x_i}), x_i||_1 \\
        &+ \frac{1}{NM} \sum_{i=1}^{N} \sum_{j=1}^{M}||G_{s^{-1}}(G_s(c_j | e_{x_i}) | e_{x_i}), c_j||_1
    \end{aligned}
    \)
     }
    & 
     \vbox{
    \( 
    \scriptsize
    \begin{aligned}
    \mathcal{L}_{idt} = & \frac{1}{NM} \sum_{i=1}^{N} \sum_{j=1}^{M} ||G_{s^{-1}}(c_j | e_{x_i}), c_j||_1  \\
    &+\frac{1}{N} \sum_{i=1}^{N} ||G_s(x_i | e_{x_i}), x_i||_1
    \end{aligned}
    \)
     } \\\bottomrule
\end{tabular}\newline
\hspace{-1.6cm}\textbf{Algorithms}\newline\newline
\scriptsize
\noindent\rule{\textwidth}{1pt}
{\scriptsize \textbf{\textsc{\footnotesize Weight Aggregation with BottleGAN}}  $K$ Clients, $G^k_{s^{-1}}$ Client Normalizing Generator, $E^k$ Client Embeddings, $G_s$ Server Restaining Generator, $E$ Server Embeddings, $T$ Communication Rounds, $e$ Number of Epochs, $D^k$ Client Datasets, $\mathcal{L}$ Loss Function }\newline\vspace{0.1cm} \noindent\rule{\textwidth}{0.4pt}
\begin{algorithmic}
\State ------ \textbf{Server executes} ---------------------------------------------------------------------\vspace{0.1cm}
\State $G^k_{s^{-1}}, E^k, G_{s}, E \leftarrow $ federated learning of \textsc{BottleGAN}
\For{each round $t=1,2,...,T$}
    \State $S_t \leftarrow$ draw randomly up to $m$ clients
    \For{each client $k \in S_t$ \textbf{in parallel}}
    \State $\hat{w}^{k}_t \leftarrow $\textsc{ClientUpdate}($k, \hat{w}_t, G^k_{s^{-1}}, E^k, G_{s}, E $)
    \EndFor
    \State $\hat{w}_{t+1} \leftarrow |K|^{-1} \sum_{k \in K} \hat{w}^k_{t}$
\EndFor
\State
\State ------ \textbf{Clients execute} --------------------------------------------------------------------\vspace{0.1cm}
\Procedure{ClientUpdate}{$k, \hat{w}_t, G^k_{s^{-1}}, E^k, G_{s}, E $}
        \State $B \leftarrow $ split $D^k$ into batches of equal size 
        \State $\hat{w}^k_t \leftarrow \hat{w}_t$
        \For{each epoch $1,2,...,e$}
            \For{$\mathcal{B} \in B$ } 
                \State $ \mathcal{B} \leftarrow G_{s^{-1}}^k(\mathcal{B} | E^k)$
                \State $\mathcal{E} \leftarrow |\mathcal{B}|$ random embeddings from $E$
                \State $ \mathcal{B} \leftarrow G_{s}(\mathcal{B} | \mathcal{E} )$
                \State $\hat{w}^k_{t} \leftarrow \hat{w}^k_{t} - \eta \nabla_{\hat{w}^k_{t}}\mathcal{L}(\mathcal{B}) $
            \EndFor
        \EndFor
    \State \Return $\hat{w}^k_t$
\EndProcedure
\end{algorithmic}\vspace{1cm}
\scriptsize
 \noindent\rule{\textwidth}{1pt}
{\scriptsize \textbf{\textsc{\footnotesize Federated Learning of BottleGAN}} $K$ Clients, $G^k_{s^{-1}}$ Client Normalizing Generator, $G^k_{s}$ Restaining Generator, $E^k$ Client Embeddings, $G_{s^{-1}}$ Server Normalizing Generator, $G_s$ Server Restaining Generator, $E$ Server Embeddings, $X^k$ Client Datasets, $C$ Destained Dataset}\newline\vspace{0.1cm} \noindent\rule{\textwidth}{0.4pt}
\begin{algorithmic}
\State ------ \textbf{Server executes} --------------------\vspace{0.1cm}
\For{each client $k=1,2,...,K$ \textbf{in parallel}}
\State $G^k_s, G^k_{s^{-1}}, E^k \leftarrow$\textsc{C-B-GAN}($k$) 
\EndFor
\State $G_s, G_{s^{-1}}, E \leftarrow$\textsc{S-B-GAN}($G^K_s, G^K_{s^{-1}}, E^K$) 
\State
\Procedure{S-B-GAN}{$G^K_s, G^K_{s^{-1}}, E^K$}
\State $\hat{X}  = \bigcup_{k=1}^K G_s^k(C | E^k) $
\State $D_s, D_c, G_s, G_{s^{-1}}, E \leftarrow$ non-federated train BottleGAN w.r.t. $\hat{X}$ and $C$

\State \Return $G_s, G_{s^{-1}}, E$
\EndProcedure
\State
\end{algorithmic}

\begin{algorithmic}
\State ------ \textbf{Clients execute} --------------------\vspace{0.1cm}
\Procedure{C-B-GAN}{$k$}
\State $D^k_s, D^k_c, G^k_s, G^k_{s^{-1}}, E^k \leftarrow$ non-federated train BottleGAN w.r.t. $X^k$ and $C$
\State \Return $G^k_s, G^k_{s^{-1}}, E^k$
\EndProcedure
\end{algorithmic}
\label{algo2}

\end{document}